\documentclass{optica-article}

\journal{opticajournal} 

\articletype{Research Article}

\usepackage{lineno}
\usepackage{subfigure}
\usepackage{color}

\usepackage{upgreek}
\usepackage[utf8]{inputenc}
\usepackage[spanish,english]{babel}

\begin{document}

\title{Intrinsic Orbital Angular Momentum Originated from Optical Catastrophe Superposition}

 \noindent\textbf{Nana Liu,\authormark{1} Huanpeng Liang,\authormark{1} Liu Tan,\authormark{1} Kaijian Chen,\authormark{1} Xiaofang Lu,\authormark{1} Shaozhou Jiang,\authormark{1} Bingsuo Zou,\authormark{1} Peilong Hong,\authormark{2,3} Jingjun Xu,\authormark{3} and Yi Liang\authormark{1,*}}

\address{\authormark{1}Guangxi Key Lab for Relativistic Astrophysics, Center on Nanoenergy Research, School of Physical Science and Technology, Guangxi University, Nanning, Guangxi 530004, China\\
\authormark{2}School of Mathematics and Physics, Anqing Normal University, Anqing, Anhui 246133, China\\
\authormark{3}The MOE Key Laboratory of Weak-Light Nonlinear Photonics, TEDA Applied Physics Institute and School of Physics, Nankai University, Tianjin 300457, China}

\email{\authormark{*}liangyi@gxu.edu.cn}

\begin{abstract*} 
Conventionally, intrinsic orbital angular momentum (OAM) is associated with phase vortices. However, our investigation into the propagation dynamics of 2D superimposed catastrophe beams, termed cyclone catastrophe beams (CCBs), reveals that these beams inherently exhibit rotation and possess OAM, distinct from the typical connection to phase vortices. Our observations clearly show these beams rotating during autofocusing propagation and particle manipulation, confirming the presence of OAM. Theoretical calculations affirm that the OAM of these beams is intrinsic and can be adjusted by varying the number of superimposed beams. Furthermore, our interference and phase studies indicate that, although CCBs exhibit phase vortices, they do not rotate around the singularities of phase vortices and their total topological charges are zero. This implies that the manifestation of OAM within CCBs does not rely on nonzero topological charge of the presented phase vortices within CCBs. Especially, eigenstates decomposition analysis illustrates that CCBs can be decomposed as a composite of Laguerre-Gaussian (LG) modes with uneven fidelity, where the topological charges of LG modes align with multiples of the superimposed catastrophe beams but do not equal to the value of the OAM per photon within CCBs, emphasizing the intrinsic OAM within CCBs and the absence of a connection to phase vortices. Our findings not only advance the understanding of the relationship between OAM and phase vortices but also pave the way for different applications of OAM waves, catalyzing their development in optics and other domains.

\end{abstract*}

\section{Introduction}
In 1992, a groundbreaking discovery by Allen illuminated that Laguerre Gaussian (LG) beams inherently carried orbital angular momentum (OAM)\cite{RN1}, with the phase center of these beams serving as a phase singularity replete with phase vortex. Building on this foundation, Allen found that nearly all phase vortices could be reinterpreted as manifestations of intrinsic OAM \cite{RN2}. This key insight sets the stage for a profound interplay between phase vortices and intrinsic OAM in subsequent developments. The presence of phase vortices was shown to inherently entail the carriage of intrinsic OAM, and conversely, the existence of intrinsic OAM inevitably gave rise to the formation of phase vortices. This intricate connection has been illustrated in many works \cite{RN3,RN4,RN5} including self-rotating beams\cite{RN279,RN281,RN7,RN6,RN8,RN282}, tornado waves\cite{RN9,RN286}, optical rotatum\cite{RN10} and azimuthal backflow\cite{RN11}. Most of these works are indeed related to the obvious modulated vortex phase or superposition of vortex beams, leading to various applications on many domains, including optical manipulation, optical communication and imaging \cite{RN3,RN4,RN5,RN284,RN12}.

Nonetheless, phase vortex and OAM represent distinct concepts, each with its unique characteristics \cite{RN13,RN14,RN16, RN287}. Phase vortices are akin to whirlpools in a fluid, exhibiting a vortex-like phenomenon where the phase surrounding them takes on a spiral wavefront \cite{RN3}. Nestled within these spirals lies a phase singularity\cite{RN15}, characterized by an uncertain phase and infinite phase gradient. OAM is rooted in the mechanics, originating from the rotational motion of an object around a center. Intriguingly, theoretical works have shown that when there are no phase vortices or the topological charge equals zero, the OAM is not necessarily zero \cite{RN16,RN287}. Their relationship only becomes apparent when the beam manifests as an eigenstate of the angular momentum operator. In sum, while phase vortices and OAM are indeed distinct phenomena, their intricate interplay have added layers of complexity and richness to optics.

Here, to gain a deeper understanding of the relationship among phase vortices and OAM, we propose and study a kind of beam rooted in catastrophe theory, named cyclone catastrophe beam (CCB). This beam originates from superimposing optical catastrophes\cite{RN17}, inherently possessing OAM even without carrying vortex phase in beam center. Experimental and theoretical investigations reveal CCBs' characteristic rotation during autofocusing propagation and their capacity to rotate particles, confirming their OAM. Furthermore, theoretical calculations unveil the intrinsic nature of CCBs' OAM, with its value decreasing as the number of superimposed beams increases. Notably, despite presenting phase vortices, the OAM per photon of CCBs deviates from the nonzero topological charge of phase vortices while the total topological charge is zero. Eigenstate decomposition analysis reveals CCBs as a composite of LG modes with asymmetric fidelity. These LG modes possess a topological charge equalling to the multiplied superimposed number. These findings offer a fresh perspective on the origin of the intrinsic OAM and significantly advances our understanding of  phase vortices, especially paving different way for OAM generation.

\section{Theoretical background}

According to catastrophe theory, a caustic field with fundamental optical catastrophe can be expressed as \cite{RN17}:  
\begin{eqnarray}
    \psi_{n}(\textbf{K})=\int_{-\infty}^{\infty}exp\left[i\phi_{n}(\textbf{K},s)\right]ds
\label{eq1}
\end{eqnarray}
where $\textbf{K}=\left({K}_{1},{K}_{2},\dots,{K}_{j}\right)$ represent the dimensionless control parameters ${K}_{j}$, 
$j=1,2,…,n-2 (n\ge3)$. $\phi_{n}\left(\textbf{K},s\right)={s}^{n}+ {\textstyle \sum_{j=1}^{j=n-2}}{K}_{j}{s}^{j}$ is the canonical potential function, $s$ is the state variable. The caustic fields in Eq. (\ref{eq1}) can present a self-accelerating propagation property \cite{RN12,RN285}. Superposing multiple such caustic fields with appropriate orthogonal dimensions and configurations, we can generate CCBs carrying OAM. For example, when we take $n=5$, an CCB in paraxial approximation can be generated by superposing 2D caustic fields with different azimuths and employing Fresnel diffraction integral:
\begin{eqnarray}
&& {\Psi}_{CCB}\left(x,y,z\right)=\frac{\exp{(i{k}z})}{i{\lambda}z}\iint_{{\mathbb{R}}}
{\textstyle\sum_{m=1}^{N}}\psi_{5}\left({K}_{m,1},0,0\right)\nonumber\\ 
&& \times{\psi}_{5}\left(0,{K}_{m,2},0\right)\exp{\left[ik\frac{(x-x')^2+(y-y')^2}{2z}\right]}d{x'}d{y'} 
\end{eqnarray}
\label{eq2}
 where $k=2{\pi}/\lambda$ is the wave number, $\lambda$ is the wavelength. ${K}_{m,j=1,2}={(-1)}^{j}\left[\cos\left(2{\pi}\frac{m-1}{N}\right)-\sin\left(2{\pi}\frac{m-1}{N}\right)\right]\frac{x}{{x}_{0}}+\left[\cos\left(2{\pi}\frac{m-1}{N}\right)+\sin\left(2{\pi}\frac{m-1}{N}\right)\right]\frac{y}{{x}_{0}}+{c}_{0}$. $N>1$ is the number of superimposed beams, ${x}_{0}$ represents transversal scale factor, ${c}_{0}$ indicates the shift of each caustic field from the center and controls the position of main lobes. 
 
\section{Results and Analysis}
\subsection{Propagation dynamics and optical manipulation}
To give a detailed analysis, we first observe the experimental propagation of CCBs by the off-axis hologram method \cite{RN18,RN19} (see detail  in Supplemental Sec. S1),  where wavelength is kept as $\lambda=632.8$ nm, input power $P=1$ W, ${x}_{0}$=47 $\mu m$, ${c}_{0}=3$. Then, we further simulate the propagation of CCBs by employing the split-step Fourier method \cite{RN18,RN19}.

\begin{figure}[htbp!]
\centering
	\includegraphics[scale=0.5]{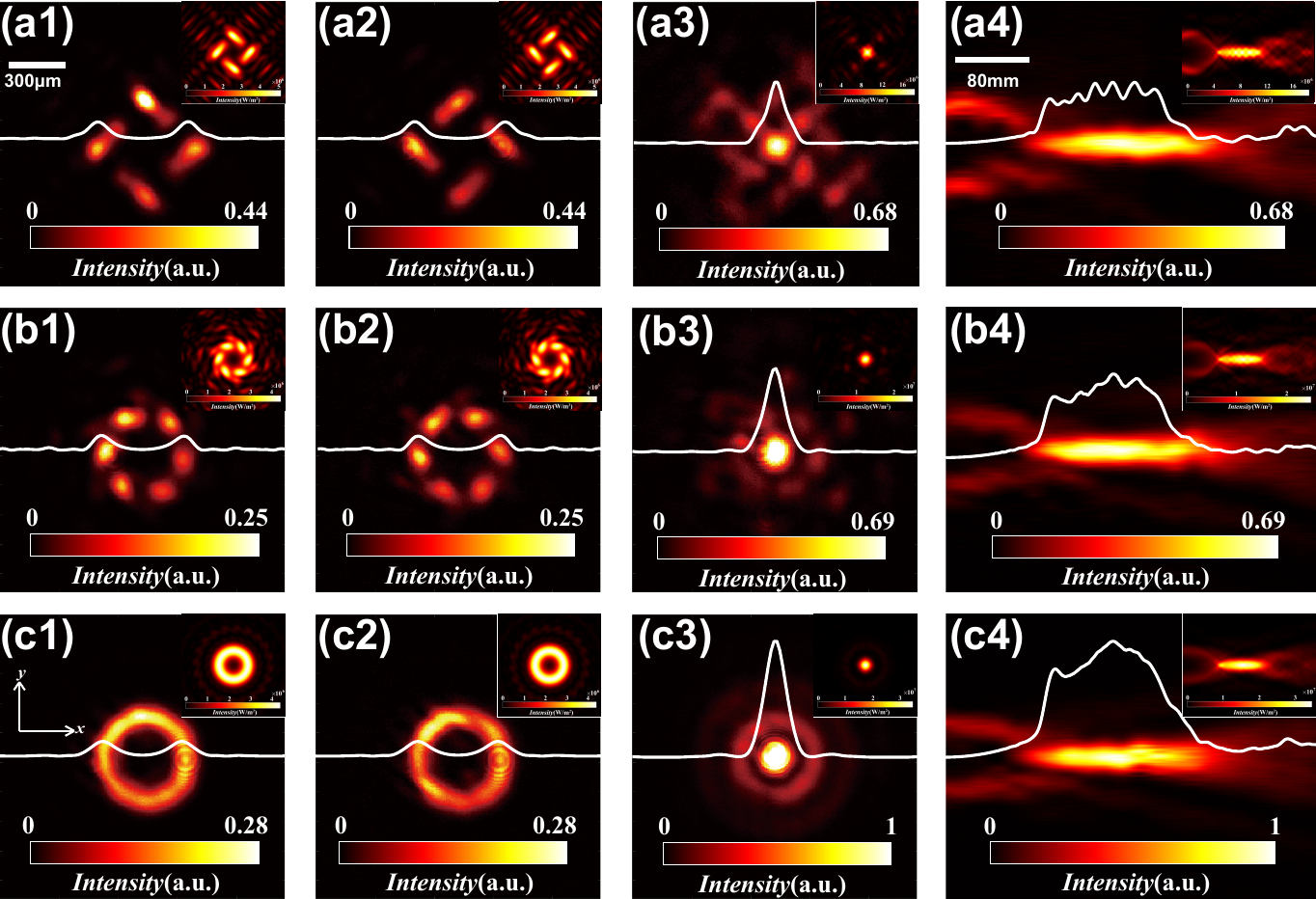}
	\caption{\label{fig:1} Propagation of CCBs with different superimposed numbers $N$. (a) $N$=4; (b) $N$=6; (c) $N$=20. (a1-c1, a2-c2) Intensity patterns of CCBs with opposite rotations at $z$=0. (a3-c3) Patterns at focus $z$=182 mm. (a4-c4)  Propagation Sideview. White lines represent the intensity profiles and insets show the simulated results.}
\end{figure}

Figure \ref{fig:1} shows the propagation dynamics of CCBs with varying superposition numbers $N$. At the source plane, the peak intensity decreases while $N$ increases [Figs. \ref{fig:1}(a1-c1)]. During propagation, CCBs undergo a rapid autofocusing process, converging into a single focus (not hollow)[Figs. \ref{fig:1}(a3-c3)]. Intriguingly, the peak intensity at focus intensifies with the increasing of the superimposed number $N$. After focus, the peak intensity of these CCBs experiences a swift decline[Figs. \ref{fig:1}(a4-c4)]. Furthermore,  Visualizations 1-6 show that, even without carrying vortex phase in beam center, the propagation of CCBs presents a clockwise rotation fashion around the beam center.  The primary lobes of CCBs gracefully arrange themselves in a circular formation, culminating in an autofocusing event at the focus. Especially, exchanging the positions of ${K}_{m,1}$ and ${K}_{m,2}$ in Eq. (\ref{eq2}), the beams experience a counter-clockwise rotation propagation process[Figs. \ref{fig:1}(a2-c2)]. All of these results indicate that CCBs possess OAM. Notably, the simulated results agree with experimental data, as evidenced by the insets within Figs. \ref{fig:1}(a-c). 

To further prove CCBs possess OAM, we use CCBs to manipulate polystyrene particles (See the detail in Supplemental Sec. S2). As shown in Figs. \ref{fig:2}(a-c) and Visualizations 7-9, driven by CCBs, these particles rotate over time. Intriguingly, the larger particle is drawn to the region of greatest intensity, securing its position in the center, while the smaller particle becomes the carrier of the larger one, guiding it along a circular path. This dynamic interaction is clearly a result of OAM within CCBs. To quantitatively scrutinize the rotational dynamics of CCBs, a white dot serves as a marker to trace the position of the smaller particle, facilitating the tracking of its rotational trajectory. Obviously, an increase in the superimposed numbers$N$ results in a gradual reduction in rotational speed . Specifically, when $N$=4, 6 and 20, the speed of rotation is 1.6, 0.9 and 0.45 rps, respectively, indicating the OAM decreases with larger superimposed number.

\begin{figure}[htbp!]
\centering
	\includegraphics[scale=0.5]{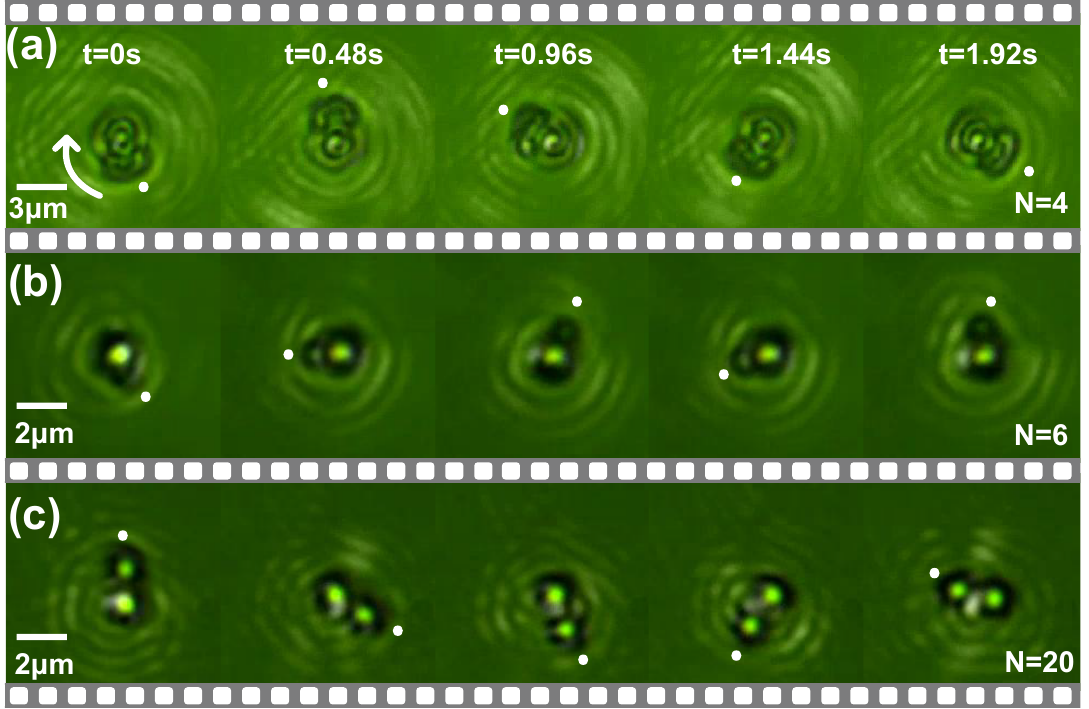}
	\caption{\label{fig:2} Polystyrene particles rotated by CCBs with different superimposed numbers $N$. (a) $N$=4; (b) $N$=6; (c) $N$=20. White dots mark the real-time position of particles.}
\end{figure}

\subsection{Analysis of OAM and phase vortices}

Generally, the time-averaged angular momentum density (AMD) in the $z$ direction can be expressed as\cite{RN20,RN21,RN22} :
\begin{eqnarray}
     {j}_z=\left(\vec{r} \times\vec{p}\right)_z=\left(\vec{r} \times\left \langle \vec{E}\times\vec{H}\right \rangle /c^2 \right)_z
\end{eqnarray}
\label{eq3}
where $\vec{p}$ is the linear momentum density, $\vec{E}$  is the electric field, $\vec{H}$ is the magnetic field, $c$ is the light speed. Then, the total OAM ${J}_{z}$ can be obtained by integrating the longitudinal AMD ${j}_{z}$ over the entire space, i.e., ${J}_{z}=\iint_{{\mathbb{R}}}{j}_zdxdy$. So, the OAM per photon $\left \langle l\right \rangle$ is\cite{RN10,RN22}:
\begin{eqnarray}
     \left \langle l\right \rangle=\frac{kc{J}_{z}}{P}=k\frac{\iint_{{\mathbb{R}}}{j}_zdxdy}{\iint_{{\mathbb{R}}}{S}_zdxdy}
\end{eqnarray}
\label{eq4}

\begin{figure}[htbp!]
\centering
	\includegraphics[scale=0.3]{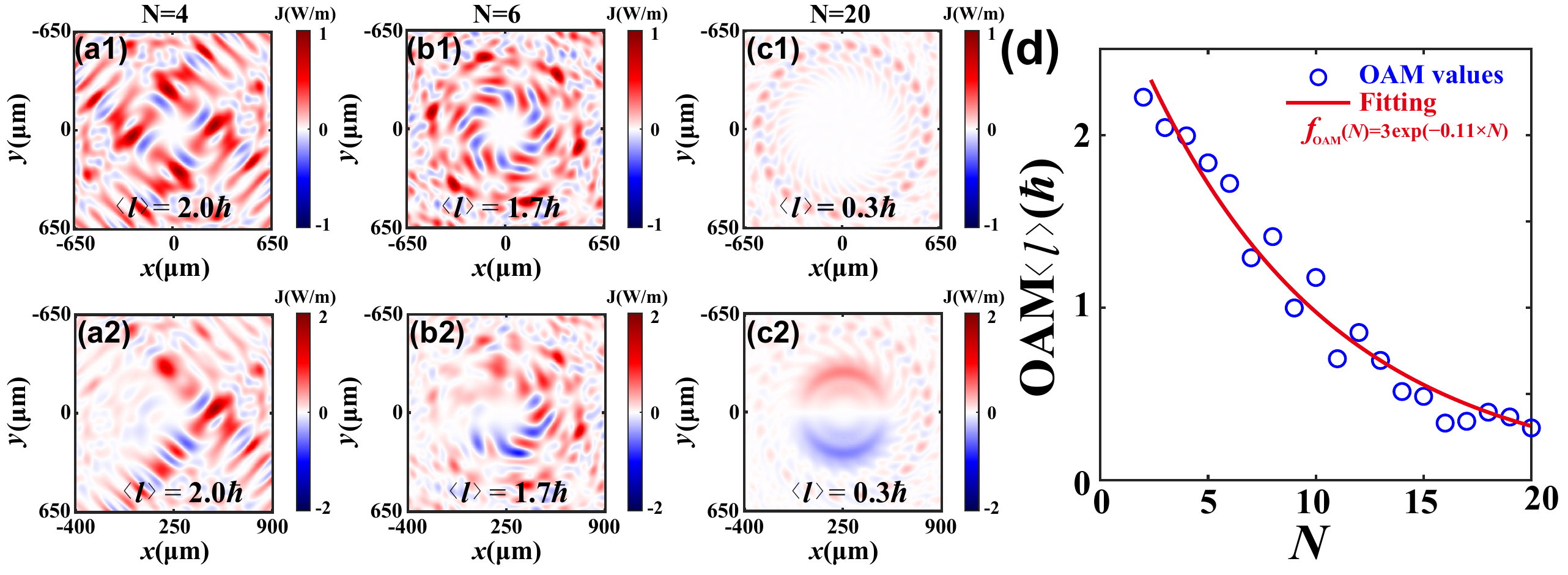}
	\caption{\label{fig:3}Longitudinal AMD distributions and OAM of CCBs at the source plane. (a-c) $N$=4, 6 and 20.  (a2-c2) AMD of CCBs after calculation axis $r$ changes. (d) Total OAM changing with $N$.}
\end{figure}

Figures \ref{fig:3}(a-c) illustrate the distribution of longitudinal AMD. As can be seen from Figs. \ref{fig:3}(a1-c1) and Visualizations 10-12, the AMD undergoes a gradual rotation as the CCBs propagate. The OAM per photon associated with CCBs is non-zero and decreases with a larger $N$. Figure \ref{fig:3}(d) qualitatively depicts the relationship between OAM and $N$, suggesting that the evolution of longitudinal OAM in CCBs can be approximately described by an exponential decay function of $N$. Furthermore, changing the calculation axis (\textit{r}) of OAM does not affect the total OAM, although the distribution changes [Figs. \ref{fig:3}(a2-c2)]. This suggests that the OAM is a type of intrinsic property, independent of the origin of  $\vec{r}$ \cite{RN20,RN21}. Actually, the total transverse linear momentums of CCBs are zero because their transverse Poynting vectors are axisymmetric (see Supplemental Sec. S3). These findings align with previous predictions, confirming the existence of intrinsic OAM in CCBs. Noted that, our calculation are analyzed in a finite room, i.e., locally. 

\begin{figure}[htbp!]
\centering
	\includegraphics[scale=0.6]{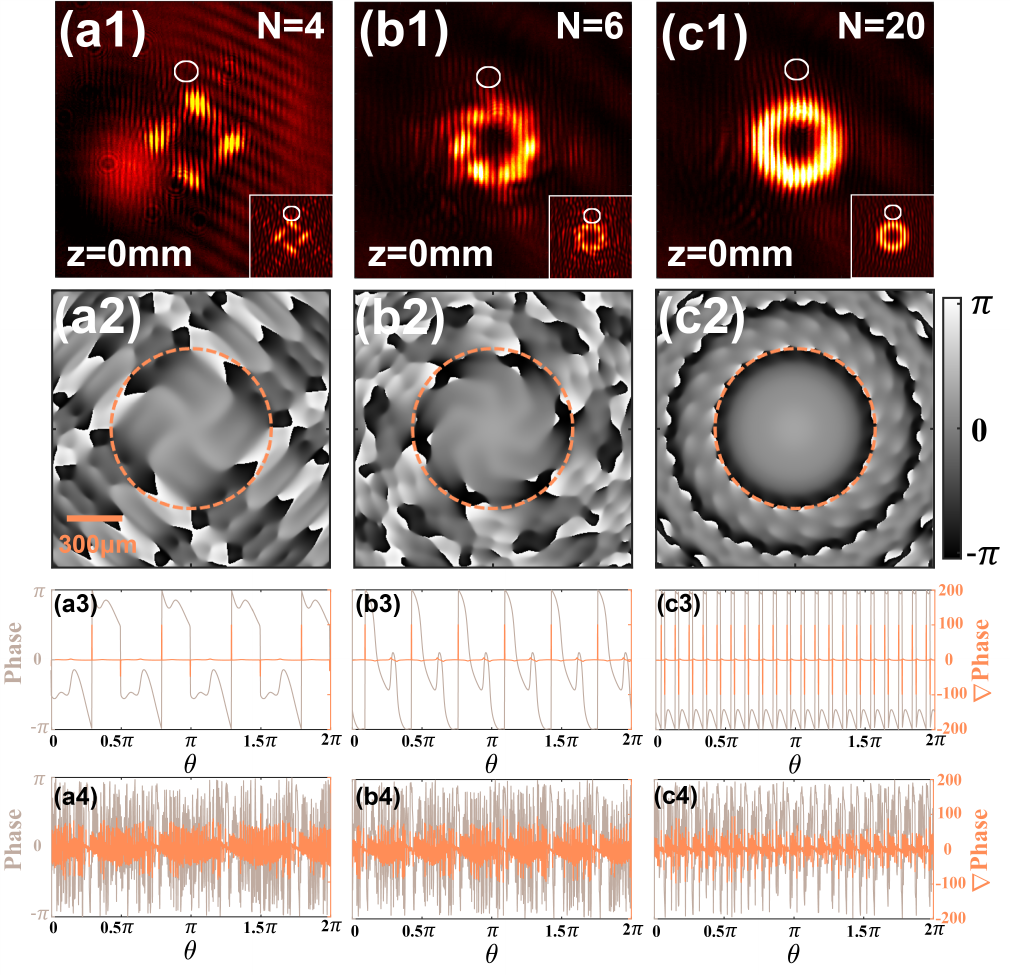}
	\caption{\label{fig:4} Interference, phase and phase gradient distributions of CCBs with varying superimposed numbers ($N$): (a-c) $N$=4, 6, and 20. (a1-c1) Interference patterns at $z$ = 0 mm. White circles mark fringe bifurcation. Insets show simulated results. (a2-c2) Phase distribution. (a3-c3, a4-c4) Phase and phase gradient profiles at $r$=425 $\mu$m (orange circles in (a2-c2)) and $r$=5 mm, respetively, varying with azimuth angle $\theta$.}
\end{figure}

As mentioned before, intrinsic OAM is usually reltated to phase vortices. The presence of a phase vortex within a plane wave can results in a splitting of the interference fringes \cite{RN3}. Therefore, we investigated the interference patterns of CCBs to find evidence of phase vortices. Figures \ref{fig:4}(a1-c1) and \ref{fig:4}(a2-c2) show the interference and phase distributions at the source plane. Clearly, while no fringe splitting occurs at the center, bifurcation appears symmetrically around other positions (marked by white circles), revealing the presence of phase vortices. Notably, each vortex only exhibits one bifurcation fringe, indicating a topological charge of 1 (2$\pi$ phase change around it). As we know, phase vortices possess  phase singularities, complementary to caustics \cite{RN287}. Usually, caustics stand out prominently in the shortwave asymptotic regime, where phase singularities are too closely packed to be clearly resolved in caustics\cite{RN287}. Here, via superposition, the interaction of caustic fields in CCBs make the caustics present their distinct features, leading to the presence of phase vortices.

To delve deeper into these phase vortices, we also calculated the total topological charge ($l_{total}$)  of CCBs. This involves summing the phase accumulated around a circle encompassing all vortices and dividing by 2$\pi$\cite{RN16}: 
\begin{eqnarray}
    l_{total}=\frac{1}{2\pi}\oint d \theta (\partial _{\theta} arg \Psi_{CCB} (r,\theta,z) )_{r{\to \infty}} .
\end{eqnarray}
\label{eq5}
At first, we calculated the phase and phase gradient at a small circle ($r$=425 $\mu$m, marked by the orange circles in Figs. \ref{fig:4}(a2-c2)).  Obviously,  phase and phase gradient exhibit periodic changes, with the phase swinging between $-\pi$ to $\pi$[Figs. \ref{fig:4}(a3-c3)]. For $N=4, 6$, the number of positive phase gradient maximum (infinity) equals to $N$, indicating that the local total topological charge is also $N$. However, for $N=20$, the total topological charge is zero due to equal positive and negative phase gradients. When we take calculation at a large circle of $r$=5 mm [Figs. \ref{fig:4}(a4-c4)], phase and phase gradient show a oscillating profiles within several wave packets (their number matches superposed number $N$), where it is difficult to distinguish the total contribution of the positive and negative phase gradients. However,  Figures \ref{fig:4}(a2-c2) show that CCBs always exhibit a multiple of $N$ charge-cancelling phase vortices, statistically resulting in a total topological charge of zero (More details about phase variations with $r$ are in Supplemental Sec. S5). Thus, this zero topological charge in Fig. \ref{fig:4} differs from the OAM per photon $\left \langle l\right \rangle$ in Fig. \ref{fig:3}. The intrinsic OAM in CCBs does not stem from these phase vortices.

For further analyzing the total topological charge and the relationship between CCBs and phase vortices, we also calculate the fidelity of projection of such beams on the LG modes ${\psi}_{l,p}$ with different topological charges $l$ via eigenstates decomposition \cite{RN22,RN23}. We consider the CCBs as a superposition of coaxial LG beams with the same waist radii, taken as unity as below:
\begin{eqnarray}
    {\Psi }_{CCB}\left(x,y,z\right)={\textstyle \sum_{p,l}^{}}{a}_{l,p}{\psi}_{l,p}(r,\theta, z)
\end{eqnarray}
\label{eq6}
where the fidelity ${a}_{l,p}$ and LG modes ${\Psi}_{l,p}$ are defined as\cite{RN22,RN23}: 
\begin{eqnarray}
    {a}_{l,p}=\frac{{\left | \left \langle {\Psi}_{CCB}  | {\psi}_{l,p}  \right \rangle \right |}^{2}}{\left \langle {\Psi}_{CCB}  | {\Psi}_{CCB}  \right \rangle \left \langle {\psi}_{l,p}  | {\psi}_{l,p}  \right \rangle }, \nonumber\\
     {\psi}_{l,p}(r,\theta, z)={N}_{l,p}r^{|l|}L_{p}^{|l|} \frac{r^2}{1+z^2} exp\left(-\frac{r^2}{2(1+iz)}\right)\nonumber\\ \times\frac{exp\left(i\left(l\theta -2p\arctan z\right)\right)}{{\left(1+iz\right)}^{|l|+1}}
\end{eqnarray}
\label{eq7}
${N}_{l,p}=\sqrt{p!\left(\pi\left(p+|l|\right)!\right)}$ and $L_{p}^{|l|}$ is the generalized Laguerre polynomial, $r=\sqrt{x^2+y^2}$, $\theta=\arctan \left(\frac{y}{x}\right)$.

\begin{figure}[htbp!]
\centering
	\includegraphics[scale=0.35]{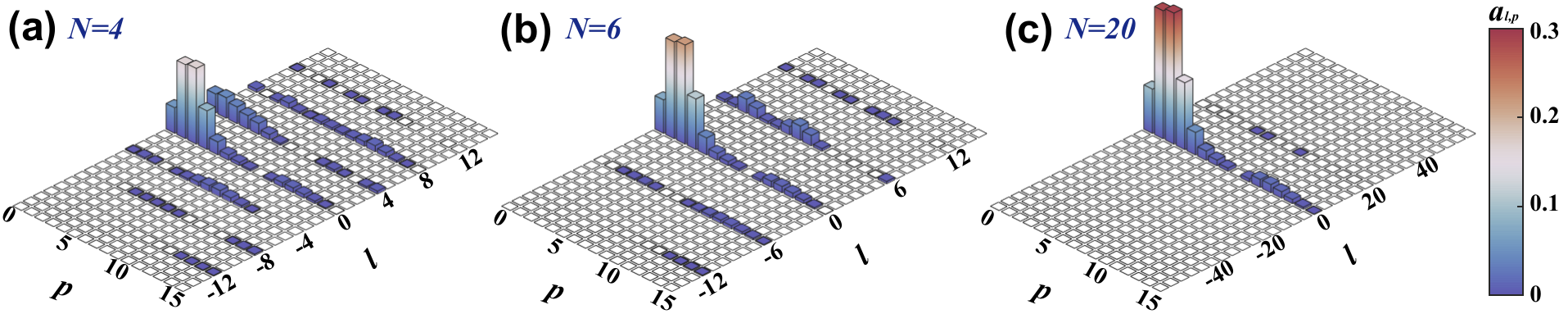}
	\caption{\label{fig:5} (a-c) Fidelity distribution of projection on Laguerre Gaussian (LG) modes ${\psi}_{l,p}$ with different topological charges $l$ in CCBs when $N$=4, 6 and 20, respectively.}
\end{figure}

Figures \ref{fig:5}(a-c) show the evolution of such fidelity of CCBs with different superimposed numbers $N$. Clearly, only the LG modes with $l=bN$($b=0, \pm1, \pm2, \dots$) possess non-zero fidelity. For a same $N$, the fidelity exhibit asymmetry distribution and an oscillation damping with larger $\left|b\right|$ and $p$, underscoring that the OAM of CCBs is not zero. When $N$ increases, the fidelity of the ${LG}_{0,p}$ modes increases and that of the ${LG}_{l=bN, p}$ modes decreases, indicating the OAM decreases with $N$. However, it should be known that, as mentioned in Ref. \cite{RN16}, the total topological charge of the beam is determined by the fidelity of LG mode with the largest value of $|l|+2p$. From Fig. \ref{fig:5}, we can infer that ${LG}_{0,p}$ modes dominates the largest contribution, i.e., the largest fidelity when the eigenstates decomposition reaches infinite room. Clearly, this finding resonates with the above consequence of zero total toplogical charge from Fig.  \ref{fig:4}. Overall, the result of LG eigenstates decomposition reaffirm the intrinsic OAM within CCBs and it is not a result of phase vortices.

\section{Dicussion and Conclusion}

Our research consistently reveals that intrinsic OAM of CCBs can be nonzero ever though the total topological charge is zero.  Actually, this local OAM originates from the inherent characteristics of the caustic in optical catastrophe. When $N$=1, these catastrophe beams present self-acceleration and exhibit both non-zero transversal linear momentum and OAM. However, the OAM in this scenario varies with calculation coordinates origin. As $N>1$, CCBs display axisymmetric-transversal linear momentum around their beam center. In this configuration, the total transversal linear momentum is zero and the OAM remains constant, irrespective of the beam’s center or coordinates. Thus, the OAM of CCBs is an intrinsic feature. Futhermore, the transversal scale factor ${x}_{0}$ dictates the acceleration of the catastrophe beams, while ${c}_{0}$ determines the position of each superposed beam from the origin of $\vec{r}$ \cite{RN17,RN285}. In this context, the OAM of CCBs increases with a larger ${c}_{0}$ and a smaller ${x}_{0}$ through Eq. (3), as elaborated in Supplemental Sec. S6. Noted that, for other CCBs, including lower or higher-order superimposed catastrophe beams ($n$=4 and 6), they exhibit similar OAM-related characteristics to the case of $n$=5, further supporting our conclusions, as detailed in Sec. S7 and S8 of the Supplement.

In summary, we've observed rotational behavior in CCBs during autofocusing propagation and optical manipulation, indicating the presence of OAM even without phase vortices in the beams center. Calculations and analysis of OAM confirm the intrinsic non-zero OAM within CCBs. Furthermore, our interference and phase studies reveal that while CCBs exhibit phase vortices, the total topological charge remains zero. These beams can be conceptualized as a superposition of Laguerre-Gaussian modes with uneven fidelity, where the topological charges of these LG modes align with multiples of the superimposed catastrophe beams. Particularly, the distinct structure of our characterized CCBs is solid at the focus, deviating significantly from conventional hollow optical vortices. This unique feature presents exciting opportunities for innovative applications in wave-matter interaction, optical manipulation, and advanced communication systems. Our work not only deepens the understanding of the relationship between phase vortices and OAM but also paves the way for OAM development. Importantly, the applicability of our findings extends beyond optical beams to OAM wave packets in various domains, including ultrasound, electron beams, and ion beams, making these structured waves versatile tools with immense potential to promote diverse sciences and technologies.


\bibliography{reference}






\end{document}